\documentclass[allclo]{FBSart}
\usepackage{amsfonts}
\usepackage{amssymb}
\usepackage{epsfig}
\usepackage{bm}
\def\Vec#1{\mbox{\boldmath $#1$}}
\def\eqne{\end{equation}}
 
\def\eqnb{\begin{equation}}

\def\NPB{{Nucl. Phys.} {\bf B}}
\def\PLB{{Phys. Lett.} B}
\def\PLA{{Phys. Lett.} A}
\def\PRL{Phys. Rev. Lett.}
\def\PRD{{Phys. Rev.} D}

\newcommand{\Slash}[1]{\ooalign{\hfil/\hfil\crcr$#1$}}

\title{The self-dual gauge fields and \\the domain wall fermion zero modes}
\author{Sadataka Furui }
\institute{School of Science and Engineering, Teikyo University.\\
1-1 Toyosatodai, Utsunomiya, 320-8551 Japan\thanks{\textit{E-mail address:} furui@umb.teikyo-u.ac.jp }}

\runningauthor{Sadataka\,Furui
 }
\runningtitle{The self-dual gauge fields and the domain wall fermion zero modes}
\begin{document}

\maketitle

\begin{abstract}
A new type of gauge fixing of the Coulomb gauge domain wall fermion system that reduces the fluctuation of the effective running coupling and the effective mass of arbitrary momentum direction including the region outside the cylinder cut region is proposed and tested in the $16^3\times 32\times 16$ gauge configurations of RBC/UKQCD collaboration.

The running coupling at the lowest momentum point does not show infrared suppression and compatible with the experimental data extracted from the JLab collaboration. The source of the fluctuation of the effective mass near momentum $p=$0.6GeV region is expected to be due to the domain wall fermion zero modes.
\end{abstract}

\section{Introduction}
The infrared (IR) QCD is characterized by the color confinement and the chiral symmetry breaking. The features are studied via Dyson - Schwinger equation and/or via lattice simulation, but the interpretation of data does not necessarily agree with each other.  We performed \cite{SF08b} lattice simulations of the domain wall fermion which preserves the chiral symmetry in the zero - mass limit using the full QCD  gauge configurations of the RBC/UKQCD collaboration of $16^3\times 32\times 16$ lattice \cite{AABB07} and compared with results of staggered fermion of the MILC collaboration\cite{MILC01}.  In these lattice simulations and in comparison with other works, we observed qualitative differences between quenched and unquenched simulations in the QCD effective running coupling, and in the  Kugo-Ojima  parameter which is calculated numerially by the inverse of the Faddeev-Popov operator at the zero momentum\cite{FN03,FN04,SF08a}. 

In \cite{FN05,FN07} we observed the running coupling of unquenched lattice simulation measured from the quark - gluon coupling and that measured from the ghost-gluon coupling are consistent, and they have enhancement as in comparison to the two loop perturbative QCD (pQCD) results in the region of the momentum $q$ around a few GeV.  The running coupling in the IR limits is suppressed in the quenched approximation \cite{FN04}. Using the unquenched configurations in the Landau gauge, we observed also a suppression, however in the Coulomb gauge \cite{FN07}, we obtained the running coupling which shows no infrared suppression and consistent with the experimental data extracted by the JLab collaboration \cite{DBCK06}. 

In the Gribov-Zwanziger Lagrangian of the ghost-gluon system, there is a term $\Delta_\gamma$ that breaks the Becchi-Rouet-Stora-Tyutin(BRST) symmetry softly,  which does not affect the renormalizability of the theory but affects the gluon propagator and the ghost propagator in the infrared \cite{DGSVV08}. In the analytical calculation, by adding a new mass term of the auxiliary field that is introduced in $\Delta_\gamma$, the gluon propagator and the ghost propagator that are consistent with lattice simulation data are discovered. In quenched lattice simulation of relatively large lattices, the infrared suppression of the Landau gauge running coupling is confirmed \cite{BIMS09}, but it is due to the absence of the infrared singularity of the ghost propagator \cite{CM08, Kondo09}. 

  The difference of the quenched and unquenched lattice simulation in the IR region could be attributed to some topological effects.  In the finite temperature Coulomb gauge QCD, van Baal \cite{vB96} studied instantons (carolons) on $T^3\times {\Vec R}$.  On the $T^3$ the Polyakov loop
$\displaystyle P_i(x)=\frac{1}{2}Tr P\,exp\{i\int_0^L A_i(x+s\hat e_i)ds\}$
is defined, and in the $A_0=0$ gauge by applying an anti-periodic gauge transformation, one could produce a non-trivial topology with the twist $P_i(t=T)=-P_i(t=0)$ and in the limit of $T\to\infty$, instanton in the $T^3\times {\Vec R}$ space was produced.  The domain walls of RBC/UKQCD collaboration are separated by the distance of $L_s$ in the 5 th dimension. In the limit of $L_s\to \infty$ the topology of the domain wall fermion and the gauge field may become diffeomorphic to $T^4\times {\Vec R}$.  Although supersymmetry is not explicitly realized on the lattice, the fermionic zero modes could affect the topology of the bosonic gauge fields, as shown in the Lagrangian field theory\cite{AdVe77}. 

 In \cite{SF08b}, we studied the domain wall fermion propagator with the cylinder cut, i.e. in the momentum region whose direction is close to that along the diagonal of the four dimensional lattice. Since the gauge configurations of RBC/UKQCD have the time axis twice as long as the space axis, we calculated the propagator of the cylinder cut which has the momentum components $(p,p,p,2p)$. In the region outside the cylinder cut, the fluctuation of the data of the effective running coupling and the effective mass were too large in the analysis using samples of the order of 100. We conjecture that the fluctuation is due to the fermionic zero mode and via a specific gauge fixing in the 5 dimensional space, based on the quaternion real condition, reduce the fluctuation.   In this paper, we extend to some general momentum configurations.

Organization of this paper is as follows. In sect.2 we present the method of choosing the gauge in the 5 dimensional space and in sect.3, we check the smallness of the deviation from the quaternion real condition, for the moment $(p_x,p_y,p_z,p_t)=( p, p, p,0), ( p,0,0,0)$ and $( p, p, p,2 p)$ with $\bar p=p a=1,2,3,4$ ($a$ is the lattice spacing). 
 The validity of the Ansatz is checked Applications to the measurement of the effective running coupling and the effective mass  are given in sect.4. Discussion and conclusion are presented in sect.5.

\section{The gauge fixing in the 5th dimension}

We specify the 5th dimension of the domain wall fermion by $s=\{0,1,\cdots,L_s-1\}$ and define the fermionic part of the Lagrangian as \cite{Sha93a, Sha93b, FurSha94}, 
\begin{equation}
S_F(\bar\psi, \psi,U)=-\sum_{x,s;y,s'}\bar\psi_{x,s}(D_F)_{x,s;y,s'}\psi_{y,s'},
\end{equation}
where
\begin{equation}
(D_F)_{x,s;y,s'}=\delta_{s,s'}D^\parallel_{x,y}+\delta_{x,y}D^\perp_{s,s'}.\label{D_F}.
\end{equation}
The action $D^\parallel$ contains the gauge field, and the interaction in the 5th dimension defined by $D^\perp$ does not contain the gauge field.

Using the $D_F$ defined in eq.\ref{D_F}
we make a hermitian operator $D_H=\gamma_5 R_5 D_F$, where $(R_5)_{ss'}=\delta_{s,L_s-1-s'}$ is a reflection operator as
\begin{eqnarray*}D_H=&&\left(\begin{array}{cccccc} -m_f\gamma_5 P_L &  & & & \gamma_5 P_R & \gamma_5(D^\parallel -1)\\& & & \gamma_5P_R & \gamma_5({D^\parallel}-1)& \gamma_5 P_L \\   & & \cdots&\cdots & \cdots&   \\ & & \cdots &\cdots &\cdots &  \\ \gamma_5P_R & \gamma_5({D^\parallel}-1)& \gamma_5 P_L & & &\\ \gamma_5({D^\parallel}-1)& \gamma_5 P_L & & & & -m_f\gamma_5 P_R \end{array}\right)\\
\end{eqnarray*}
where $P_{R/L}=(1\pm\gamma_5)/2$, and -1 in $(D^\parallel -1)$ originates from $D^\perp_{s,s'}$.

The quark sources are sitting on the domain walls as
\begin{equation}
q(x)=P_L\Psi(x,0)+P_R\Psi(x,L_s-1).
\end{equation}
We take $P_L\Psi(x,s)\propto e^{-(\frac{s}{r})^2}$, $P_R\Psi(x,s)\propto e^{-(\frac{L_s-s-1}{r})^2}$, $(s=0,1, \cdots, L_s-1)$ with $r=0.842105$, which corresponds to $(r/L_s)/(1+r/L_s)=0.05$.

The spinor basis that we adopt for the domain wall fermion is $\gamma_5$ diagonal i.e. 
\[
\gamma_\mu=\left(\begin{array}{cc} 0 & \sigma_\mu\\
                            (\sigma_\mu)^\dagger& 0\end{array}
\right),\qquad
\gamma_5=\left(\begin{array}{cc} I &0\\
                          0 & -I\end{array}
\right)
\]
where $\displaystyle \gamma_0=\left(\begin{array}{cc} 0 & I\\
                            I& 0\end{array}\right)$ and $\sigma_i$ (i=1,2,3) are the Pauli matrices
\begin{equation}
\sigma_1=\left(\begin{array}{cc}0 & i\\
                          i & 0\end{array}\right),\qquad
\sigma_2=\left(\begin{array}{cc}0 & -1\\
                          1 & 0\end{array}\right),\qquad
\sigma_3=\left(\begin{array}{cc}i & 0\\
                          0 & -i\end{array}\right).
\end{equation}

The domain wall fermion propagator of the momentum $p$ can be expressed as
\begin{equation}
S(p)=[\frac{-i\Slash{p}+{\mathcal M}^\dagger(\hat p)}{p^2+{\mathcal M}(\hat p){\mathcal M}^\dagger(\hat p)}P_L]+[\frac{-i\Slash{p}+{\mathcal M}(\hat p)}{p^2+{\mathcal M}^\dagger(\hat p){\mathcal M}(\hat p)}P_R]
\end{equation}
where using the plane wave with the coordinate of the 5th dimension $s$ specified as $\chi(p,s)$ and the solution of the conjugate gradient method $\Psi(p,s)$, overlap Tr$\langle \bar \chi(p,s)P_{L/R}\Psi(p,s)\rangle\propto {\mathcal B}_{L/R}$ and Tr$\langle \bar \chi(p,s)i\Slash{p} P_{L/R}\Psi(p,s)\rangle\propto{\mathcal A}_{L/R}$, the mass is defined as
\[
{\mathcal M}(\hat p)=\frac{Re[{\mathcal B}_R(p,L_s/2)]}{Re[{\mathcal A}_R(p,L_s/2)]}
\]
and \[
{\mathcal M}^\dagger(\hat p)=\frac{Re[{\mathcal B}_L(p,L_s/2)]}{Re[{\mathcal A}_L(p,L_s/2)]}.
\]

Here, ${\mathcal M}^\dagger (\hat p){\mathcal M}(\hat p)$ has a zero mode, while ${\mathcal M}(\hat p){\mathcal M}^\dagger(\hat p)$ has no zero mode\cite{NN93}. 
In \cite{SF08b}, we studied the domain wall fermion propagator in the Coulomb gauge adopting the cylinder cut, i.e. diagonal directions in the 4-d momentum space $n_x\times n_y\times n_z\times n_t$ spacial extension are selected. In this paper we consider other momenta.

\subsection{The U(1) real condition}
 Since we adopt the representation of fermion spinors such that the $\gamma_5$ is diagonal, the fermion field $\psi$ can be gauge transformed in the 5th dimension as
\begin{equation}
\psi\to e^{i\eta\gamma_5}\psi, \qquad \bar\psi \to \bar\psi e^{-i\eta\gamma 5},
\end{equation}
such that on the left and on the right domain wall, the propagator becomes approximately real by choosing\cite{SF08b}
\[
|e^{i\theta L}e^{i\eta}-1|^2+|e^{i\theta R}e^{-i\eta}-1|^2
\]
be minimum ($a$-type), or
\[
|e^{i\theta L}e^{i\eta}+1|^2+|e^{i\theta R}e^{-i\eta}-1|^2
\]
be minimum ($b$-type), where we define
\begin{eqnarray}
e^{i\theta_L}&=&\frac{Tr\langle \chi(p,0)\phi_L(p,0)\rangle}{|Tr\langle \chi(p,0)\phi_L(p,0)|}\nonumber\\
e^{i\theta_R}&=&\frac{Tr\langle \chi(p,L_s-1)\phi_R(p,L_s-1)\rangle}{|Tr\langle \chi(p,L_s-1)\phi_R(p,L_s-1)|}.
\end{eqnarray}
Here, $Tr$ is the trace in the color-spin space. 
\begin{eqnarray}
&&Tr\langle \widetilde{\chi(p_x,0/L_s-1)\phi_{L/R}(p_x,0/L_s-1)}\rangle \sigma_1+Tr\langle\widetilde{\chi(p_y,0/L_s-1)\phi_{L/R}(p_y,0/L_s-1)}\rangle\sigma_2\nonumber\\
&&+Tr\langle \widetilde{\chi(p_z,0/L_s-1)\phi_{L/R}(p_z,0/L_s-1)}\rangle \sigma_3
+Tr\langle \widetilde{\chi(p_z,0/L_s-1)\phi_{L/R}(p_z,0/L_s-1)}\rangle i I
\nonumber\\
&&=({a_x}^{0/L_s-1} \sigma_1+{a_y}^{0/L_s-1} \sigma_2+{a_z}^{0/L_s-1} \sigma_3+{a_t}^{0/L_s-1} i I)\sqrt{det A^{L/R}}
\end{eqnarray}
where 
\[
det A^{L/R}=det({a_x}^{0/L_s-1} \sigma_1+{a_y}^{0/L_s-1} \sigma_2+{a_z}^{0/L_s-1} \sigma_3+{a_t}^{0/L_s-1}iI).
\]

The spinor of the fermion on the domain walls can be expressed in quaternion bases $H$ which is obtained by adjoining an element $j$ to the complex field $C$,
i.e. $H=C+Cj\simeq C^2$ \cite{AW77}.  The equation in 4 dimensional space is converted in projective 3 dimensional space $P_3$ and the antilinear map $\sigma: P_3\to P_3$ with $\sigma^2=1$ defines a real structure on $P_3$.

 The instanton solution proposed by Atiyah et al. \cite{AHDM78} is produced in the quaternion bases and the transformation matrix which satisfies a specific condition which is called quaternion real condition was crucial for establising the self-duality \cite{CG81}. 
We study whether the domain wall fermion on the left boundary can be transformed to that on the right boundary by this type of transformation matrix. 

In the lattice simulation of the domain wall fermion \cite{SF08b} we sample wise measure the propagator by using 
the transformation matrix from the left boundary
\[
a_x^0\sigma_1+a_y^0\sigma_2+a_z^0\sigma_3+a_t^0\,i I=\left(\begin{array}{cc}a_{0}& b_{0}\\
                       c_{0}& d_{0}\end{array}\right)
\]
which is given by the sample average of color diagonal components 
to the right boundary
\[
a_x^{L_s-1}\sigma_1+a_y^{L_s-1}\sigma_2+a_z^{L_s-1}\sigma_3+a_t^{L_s-1}\,i I=\left(\begin{array}{cc}a_{L_s-1}& b_{L_s-1}\\
                       c_{L_s-1}& d_{L_s-1}\end{array}\right)
\]

\subsection{The quaternion real condition}

In the analysis of instantons in quaternion bases, Corrigan and Goddard \cite{CG81} defined the transition function $g(\omega,\pi)$ where  $\pi$ is the complex two component spinor  which satisfy.
\[
g(\lambda\omega,\lambda\pi)=g(\omega,\pi), \quad det\, g=1.
\]
Here $\lambda$ is s non-zero complex constant, $\omega=p\pi$ where $p$ is a quaternion and the transformation $g(p\pi,\pi)$ between the domain walls should satisfy the reality condition\cite{CG81}. In this case, a specific Ansatz can be introduced, which is discussed in sect.3.

@When $\omega=p\pi$ where $p=p^0-i {\Vec p}\cdot{\Vec\sigma}$, $g(\omega,\pi)$ is a quaternion can be expressed as
\[
g(p\pi,\pi)=h(p,\zeta)k(p,\zeta)^{-1}
\]
where $\zeta=\frac{\pi_1}{\pi_2}$, $h(p,\zeta)$ is regular in $|\zeta|>1-\epsilon$ and $k(p,\zeta)$ is regular in $|\zeta|<1+\epsilon$.

In \cite{CG81},  the Ansatz
\begin{eqnarray}
g_0&=&\left(\begin{array}{cc}e^{-\nu}&0\\
                           0&e^\nu\end{array}\right)
\left(\begin{array}{cc}\zeta^1&\rho\\
                       0&\zeta^{-1}\end{array}\right)
\left(\begin{array}{cc}e^{\mu}&0\\
                           0&e^{-\mu}\end{array}\right)\nonumber\\
&=&\left(\begin{array}{cc}e^\gamma\zeta^1&f(\gamma,\zeta)\\
                       0&e^{-\gamma}\zeta^{-1}\end{array}\right)
\end{eqnarray}
was proposed as the transformation matrix.  In our 5-dimesional domain wall fermion case, $\gamma=\mu-\nu$ and $\mu, \nu$ contain the phase in the 5th direction $i\eta$. 
\begin{equation}
2\mu=i\omega_2/\pi_2-i\eta=(p_x+ip_y)\zeta+ip_t-p_z-i\eta
\end{equation}
\begin{equation}
2\nu=i\omega_1/\pi_1+i\eta=(p_1x-ip_y)\zeta+ip_t+p_z+i\eta
\end{equation}
 The quaternion reality condition of the transformation matix $g(\gamma,\zeta)$ gives
\begin{equation}
\left(\begin{array}{cc}a_{L_s-1}& b_{L_s-1}\\
                       c_{L_s-1}& d_{L_s-1}\end{array}\right)
\left(\begin{array}{cc}\zeta^{1}e^\gamma& f\\
                       0&\zeta^{-1}e^{-\gamma}\end{array}\right)
=\left(\begin{array}{cc}\zeta^{1}e^{-\gamma}& \bar f\\
                       0&\zeta^{-1}e^{\gamma}\end{array}\right)
\left(\begin{array}{cc}a_{0}& b_{0}\\
                       c_{0}& d_{0}\end{array}\right)
\end{equation}
where $\displaystyle \bar f=\overline{f(\bar\gamma,-\frac{1}{\bar\zeta})}$.

The function $f$ and $\bar f$ taken in \cite{CG81} is.
\[
f=\frac{d_0 e^\gamma-\frac{1}{a_{L_s-1}}e^{-\gamma}}{\psi},\quad 
\bar f=\frac{\frac{1}{d_{L_s-1}} e^\gamma-a_0e^{-\gamma}}{\psi}.
\]

In general $c_0$ and $c_{L_s-1}$ are polynomials of $\zeta$, and they satisfy $c_{L_s-1}\zeta^{1}=c_0 \zeta^{-1}$.  We define.
\[
\psi=\hat c_{-1}\zeta^{-1}+\hat c_{1}\zeta^{1}+\delta
\]
where $\hat c_{-1}=c_{L_s-1}$ and $\hat c_{1}=c_0$ and $\delta$ is a constant, which is defined later. 

The difference
\begin{equation}
\Delta L/R=\left(\begin{array}{cc}a_{L_s-1}& b_{L_s-1}\\
                       c_{L_s-1}& d_{L_s-1}\end{array}\right)_{L/R}
\left(\begin{array}{cc}\zeta^{1}e^\gamma& f\\
                       0&\zeta^{-1}e^{-\gamma}\end{array}\right)
-\left(\begin{array}{cc}\zeta^{1}e^{-\gamma}& \bar f\\
                       0&\zeta^{-1}e^{\gamma}\end{array}\right)
\left(\begin{array}{cc}a_{0}& b_{0}\\
                       c_{0}& d_{0}\end{array}\right)_{L/R}
\end{equation}
should be small, if the left wall and the right wall are correlated by the self-dual gauge transformation.
With the Ansatz $\displaystyle \psi=c_0\zeta^{-1}+c_{L_s-1}\zeta^{1}+\delta$, we find $\displaystyle \zeta^{1}=\sqrt{\frac{c_0}{c_{L_s-1}}}$ and 
the $e^\gamma$ and the $\delta$ are calculated from the simultaneous equation (\ref{gamma}).
\begin{eqnarray}
\left\{\begin{array}{l}
a_{L_s-1}f+b_{L_s-1}\zeta^{-1}e^{-\gamma}=b_0\zeta^{-1}e^{-\gamma}+d_0 \bar f\nonumber\\
c_{L_s-1}f+d_{L_s-1}\zeta^{-1}e^{-\gamma}=d_0\zeta^{-1}e^{\gamma}.
\end{array}\right.\qquad\qquad\qquad\qquad  (14)\label{gamma}
\end{eqnarray}
The equation has two sets of solutions, which can be obtained numerically, using Mathematica\cite{Math}.

Since $g(\gamma,\zeta) A_0 g^{-1}(\gamma,\zeta) =A_{L_s-1}$ and since the factor $e^\gamma$ depends on the boundary conditions, there appears slight difference in the $a$-type minimization and the $b$-type minimization. We found globally $a$-type gives smaller deviation than the $b$-type, and we adopt the $a$-type in this paper.
From the numerical practice, we observe that the deviation $\Delta L/R$ becomes small when the $|\delta|$ is large, i.e. when $\psi\sim\delta\sim \frac{1}{f}$ is large. It means that $\displaystyle e^{2\gamma}\sim\frac{d_{L_s-1}}{d_0}$ is a good approximation in the case of small deviation.

\section{Numerical check of the deviation from the reality condition}
We calculated the quark propagator from 149 samples of the full QCD gauge configurations DWF$_{01}$ produced by the RBC/UKQCD collaboration \cite{AABB07}. We first fixed the gauge to the Landau gauge and then to the minimum Coulomb gauge. We did not perform the remnant gauge fixing of $A_0$, since we do not want to affect the topological structure of the $A_0$. The propagator calculation was done by using the conjugate gradient method \cite{SF08b}. 
We checked the deviation of the transformation matrix using the phase $\eta$ of $a$-type U(1) real condition and parameter $\delta$ and $\zeta$ of quaternion real condition,  for the momentum $(p_x,p_y,p_z,p_t)=(p, p,p,2 p)$ i.e. cylinder cut, $( p, p, p,0)$ i.e. no energy transfer and $( p,0,0,0)$ i.e. momentum transfer along the x-axis, using 49 samples of DWF$_{01}$ configuration.

We choose $\zeta^{1}=\sqrt{\frac{c_0}{c_{L_s-1}}}$ to make $(\Delta L/R)_{21}$ consistent with 0 and obtain $e^\gamma$ and $\delta$, which approximately satisfy the quaternion reality. The good indicator of the reality condition is $(\Delta L/R)_{12}$. In the tables, relatively large mean value and the same order of magnitude of the standard deviation means presence of exceptional samples.

The results for the momenta in cylinder cut are shown in Table\,\ref{dLRb}.  
 The deviation in the case of ${\Vec p}=(p, p, p,0)$ is shown in Table\,\ref{dLR} and that for ${\Vec p}=(p,0,0,0)$ is shown in Table\,\ref{dLRa}.
\begin{table}[htb]
\begin{center}
\begin{tabular}{ccccccc}
  $\bar p$ & L/R & $(\Delta L/R)_{11}$  &$(\Delta L/R)_{12}$ &$(\Delta L/R)_{21}$ &$(\Delta L/R)_{22}$&   \\
\hline
 0 &  L & 2.6(9.9)$\times 10^{-8}$ &1.38(0.89)$\times 10^{-4}$& 0 & 1.14(2.18)$\times 10^{-7}$ \\
   &  R & 3.4(10.4)$\times 10^{-8}$ &1.14(0.07)$\times 10^{-4}$&0 & 1.67(1.65)$\times 10^{-7}$\\
 1 &  L & 2.4(4.2)$\times 10^{-2}$ &0.77(1.56) &0& 0.17(0.48) \\
   &  R & 0.204(0.405) &0.546(0.523) &0 &0.062(0.114) \\
 2 &  L & 0.302(0.611) &0.87(1.01) &0& 0.30(0.69) \\
   &  R & 0.98(4.61) &1.59(5.81) &0& 0.96(4.93) \\
 3 &  L & 0.72(1.62) &1.26(1.94) &0& 0.65(1.18) \\
   &  R & 0.87(2.47) &1.36(2.13) &0& 0.74(2.17) \\
 4 &  L & 1.55(3.17) &1.72(2.23) &0& 1.25(2.92) \\
   &  R & 0.51(1.20) &0.91(1.28) &0& 0.54(1.61) \\
\hline
\end{tabular}
\caption{The deviation of the absolute value of the components of the transformation matrix for ${\Vec p}=(p, p, p, 2p)$. Numbers in the brackets are the standard deviations.} \label{dLRb}
\end{center}
\end{table}

\begin{table}[htb]
\begin{center}
\begin{tabular}{ccccccc}
  $\bar p$ & L/R & $(\Delta L/R)_{11}$  &$(\Delta L/R)_{12}$ &$(\Delta L/R)_{21}$ &$(\Delta L/R)_{22}$&   \\
\hline
 1 &  L & 4.3(1.4)$\times 10^{-3}$ &1.6(1.3) &0& 0.13(0.02) \\
   &  R & 5.9(1.1)$\times 10^{-3}$ &0.117(0.007) &0 &3.97(0.06)$\times 10^{-3}$ \\
 2 &  L & 0.62(1.47) &1.44(1.89) &0& 0.56(1.08) \\
   &  R & 2.05(1.84) &1.42(1.10) &0& 0.63(0.81) \\
 3 &  L & 2.53(9.24) &4.0(12.5) &0& 2.56(7.51) \\
   &  R & 0.52(0.52) &0.73(0.61) &0& 0.58(0.47) \\
 4 &  L & 1.52(2.90) &1.80(2.82) &0& 1.64(2.35) \\
   &  R & 2.56(4.49) &1.95(3.31) &0& 1.79(4.14) \\
\hline
\end{tabular}
\caption{The deviation of the absolute value of the components of the transformation matrix for ${\Vec p}=(\bar p,\bar p,\bar p,0)$. Numbers in the brackets are the standard deviations.} \label{dLR}
\end{center}
\end{table}

\begin{table}[htb]
\begin{center}
\begin{tabular}{ccccccc}
  $\bar p$ & L/R & $(\Delta L/R)_{11}$  &$(\Delta L/R)_{12}$ &$(\Delta L/R)_{21}$ &$(\Delta L/R)_{22}$&   \\
\hline
 1 &  L & 0.214(0.314) &7.18(4.68) &0& 1.11(0.94) \\
   &  R & 0.145(0.877) &0.217(0.238) &0 &0.08(0.47) \\
 2 &  L & 0.276(0.674) &1.80(5.85) &0& 0.36(0.84) \\
   &  R & 4.67(4.77) &1.49(1.17) &0 & 0.42(0.70) \\
 3 &  L & 0.303(0.441) &12.1(7.0) &0& 2.2(1.3) \\
   &  R & 0.027(0.011) &0.24(0.06) &0 & 0.017(0.009) \\
 4 &  L & 2.55(8.02) &2.98(5.59) &0& 2.39(6.44) \\
   &  R & 3.84(8.07) &1.84(5.26) &0& 2.18(5.99) \\
\hline
\end{tabular}
\caption{The deviation of the absolute value of the components of the transformation matrix for ${\Vec p}=( p,0,0,0)$. Numbers in the brackets are the standard deviations.} \label{dLRa}
\end{center}
\end{table}

In the case of ${\Vec p}=(4/a,0,0,0)$ we observe an exceptional sample which has a large deviation, but we omitted this sample from the average. 

\section{The effective running coupling and the effective mass of the quark}
Corresponding to the sample-wise selection of larger absolute value of $\delta$, i.e. large zero-mode component, we assign a parameter $ind=1$ for larger and 2 for smaller, and multiply the phase $(-1)^{ind}$ to the expectation value of the quark wave function that is used in the calculation of the effective mass of the quark. 
This phase factor can be absorbed in the phase $e^{i\eta}$ introduced to approximately satisfy the U(1) real condition.

We expect this phase ambiguity comes from the finite size of the length of the 5th dimension, and this phase reduces the fluctuation dramatically. The mass function as a function of $p$(GeV) is shown in Figure 1.
\begin{figure}[htb]
\begin{center}
\includegraphics[width=7cm,angle=0,clip]{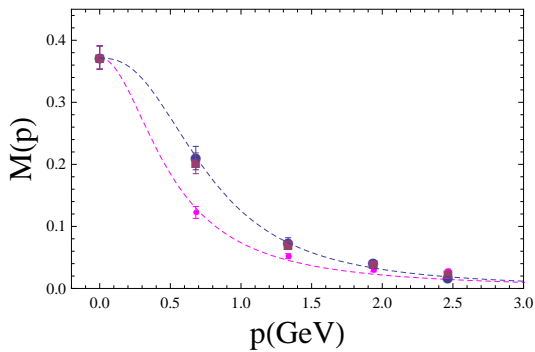}
\caption{The effective mass of the domain wall fermion. $m_u=0.01/a, m_s=0.04/a$, without reality selection (upper dashed line) and with reality selection(lower dashed line). }\label{m5enn0}
\end{center}
\end{figure}

The fitted parameters of the mass function of the DWF and the staggered fermion are given in Table 2. 
\begin{table}[htb]
\begin{center}
\begin{tabular}{cccccc}
 &$m_{ud}/a$ & $m_s/a$ & $c$ &  $\Lambda$(GeV) & $\alpha$ \\
\hline
DWF$_{01}$ &0.01   &0.04 &  0.49  &  0.76(2)  & 1.25 \\
DWF$_{02}$ &0.02  &0.04  &  0.48  &  0.80(3)  & 1.25 \\
DWF$_{03}$ &0.03  &0.04  &  0.61  &  0.66(2)  & 1.25 \\
\hline
MILC$_{f1}$ &0.006 &0.031  &  0.45 & 0.82(2) & 1.00 \\
MILC$_{f2}$ &0.012 &0.031  &  0.43 & 0.89(2) & 1.00 \\
\hline
\end{tabular}
\end{center}
\caption{The fitted parameters of mass function of DWF(RBC/UKQCD) and staggered fermion (MILC).}\label{massfunc}
\end{table}
In \cite{Adelaide01}, the self-dual gauge configuration with $|Q|=1$ had large fluctuation on $T^4$ configuration, and it was claimed that the Nahm transformation does not produce $|Q|=1$ configuration in the $T^4$ system. The 5 dimensional space of the domainwall does not have periodicity in the 5th dimension and the instanton like object that relates the fermion on the left wall and the right wall are not produced by the Nahm transformation on the fermionic zero mode. The self-dual configuration constructed \`a la ADHM and the Ansatz of Colligan and Goddard for establishing the quarternion reality was used in performing the gauge fixing of the domain wall fermion. The additional phase that is introduced following the sample dependent size of the zero-mode allows the instanton-like effects to become manifest. 

As shown in Figure 2, the QCD effective coupling of DWF in cylinder cut is slightly larger than that of the staggered fermion (dashed line), but after applying the reality condition, it becomes slightly smaller. 
An enhancement in the region of $q\sim$ a few GeV is expected to be due to $A^2$ condensates\cite{Orsay02,KMSI02}.
 Although it is necessary to study the continuum limit, the simulation of this size suggests  a large part of the fermion configurations on the two walls are correlated by the self-dual gauge fields which plays the same role as instantons.

\begin{figure}[htb]
\begin{center}
\includegraphics[width=7cm,angle=0,clip]{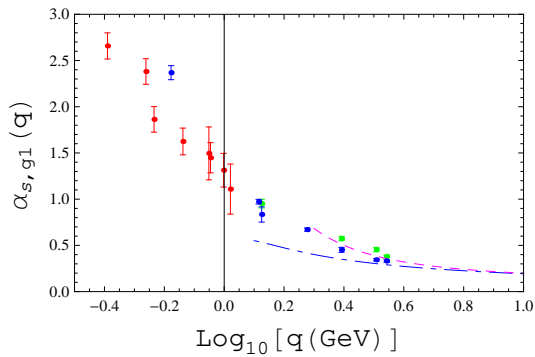}
\caption{The effective coupling $\alpha_{s,g_1}(q)$ of domain wall fermion and the gluon without reality selection (greenjand with reality correction(magenta)  in cylinder cut region. Dark low momentum points are that with reality correction.  Below 1.2GeV, red points are the experimental data of the JLab collaboration and the dark point is our result of momenta $q=0.7$GeV/c. The dash-dotted line is the two loop pQCD result and the dashed line is the pQCD with $A^2$ condensates efffect derived from the fitting the running coupling \cite{FN05} calculated by using the gauge configuration of full-QCD staggered fermion of the MILC collaboration \cite{MILC01}. 
 }\label{alp_ann}
\end{center}
\end{figure}

As was shown in \cite{SF08b}, the effective coupling in the IR region is experimentally investigated from the spin dependent Bjorken sum rule at large momenta and Drell-Hearn-Gerasimov sum rule at low momenta \cite{DBCK06}. In the literature there is also a Bethe-Salpeter analysis of the meson spectra \cite{meson07}. Our data is consistent with the JLab data and suggest saturation of the coupling constant to a finite value and inconsistent with the meson data which suggest infrared suppression.  The meson data at the lowest momentum point are based on transitions at high angular momentum states and rather ambiguous\cite{Sh08}. 

\section{Discussion and Conclusion}

The fermionic field on the domain wall at $s = 0$ can be transformed to that on  $s = L_s - 1$ by a gauge transformation whose deviation from the quaternion reality  $\Delta L/R$ is small. 
 The quaternion real condition fixes parameters $e^\gamma, \zeta$ and $\delta$ and we distinguish the two sets by the absolute value of $\delta$, i.e.
the parameter characterizing the zero-mode. One solution with large absolute value of the zero mode yields smaller $\Delta L/R$. 
 This selection reduces fluctuation of the effective mass of the quark and other physical quantities derived from IR - QCD domain wall fermion lattice simulation.

 It remains to check the finite size effects by extending the analysis to larger lattices\cite{RBC08}. It might be interesting to check how large the deviation from the reality condition $\Delta L/R$ is in the quenched configurations.

Extension of the Nahm transformation which was done in the $T^3\times {\Vec R}$ configuration \cite{vB96} to the $T^4\times {\Vec R}$ and to study the instanton would be an interesting subject. 
Although there remains an explicit calculation of the field of the instantons, we have observed that there is an evidence of the self-dual gauge field contribution in the domain wall fermion propagator. 

\vskip 0.2 true cm
\begin{acknowledge}
The numerical simulation was performed on Hitachi-SR11000 at High Energy Accelerator Research Organization(KEK) under a support of its Large Scale Simulation Program (No.07-04 and No.08-01), and on NEC-SX8 at Yukawa institute of theoretical physics of Kyoto University and at RCNP/CMC of Osaka university.
\end{acknowledge}
\end{document}